\documentclass[reprint,superscriptaddress,aps,pre,footinbib,showkeys,showpacs]{revtex4-1}

\usepackage[english]{babel}
\usepackage[T1]{fontenc}

\usepackage{hyperref}
\usepackage{amsmath}%
\usepackage{amsfonts}%
\usepackage{amssymb}%
\usepackage[usenames,dvipsnames]{color}%
\usepackage{graphicx}

\begin{document}

\title{Enhancing resilience of interdependent networks by healing}
\author{Marcell Stippinger}
\email{stippinger@phy.bme.hu}
\affiliation{Department of Theoretical Physics, 
Budapest University of Technology and Economics, \\
H-1111 Budafoki \'ut 8., Budapest, Hungary}
\author{J\'anos Kert\'esz}
\email{kertesz@phy.bme.hu}
\affiliation{Center for Network Science,
Central European University,
H-1051 N\'ador u. 9., Budapest, Hungary}
\affiliation{Department of Theoretical Physics, 
Budapest University of Technology and Economics, \\
H-1111 Budafoki \'ut 8., Budapest, Hungary}

\begin{abstract}
Interdependent networks are characterized by two kinds of interactions:
The usual connectivity links within each network and the dependency links
coupling nodes of different networks. Due to the latter links such networks
are known to suffer from cascading failures and
catastrophic breakdowns. 
When modeling these phenomena, usually one assumes that a fraction of nodes
gets damaged in one of the networks, which is followed possibly by a
cascade of failures. In real life the initiating failures do not occur at
once and effort is made replace the ties eliminated due to the failing
nodes.
Here we study a dynamic extension of the model of interdependent networks
and introduce the possibility of link formation with a probability $w$,
called healing, to bridge non-functioning nodes and enhance network
resilience.
A single random node is removed, which may initiate an avalanche. After
each removal step healing sets in resulting in a new topology. Then a new
node fails and the process continues until the giant component disappears
either in a catastrophic breakdown or in a smooth transition.
Simulation results are presented for square lattices as starting networks
under random attacks of constant intensity.
We find that the shift in the position of the breakdown
has a power-law scaling as a function of the healing probability with
an exponent close to~$1$.
Below a critical healing probability, catastrophic cascades form and
the average degree of surviving nodes decreases monotonically,
while above this value
there are no macroscopic cascades and the average degree has first an
increasing character and decreases only at the very late stage of the
process.
These findings facilitate to plan intervention in case of crisis situation
by describing the efficiency of healing efforts needed to suppress cascading
failures.
\end{abstract}

\maketitle

\section{Introduction}
\label{sec:introduction}

Robustness is one of the key issues for network maintenance and
design~\cite{Cohen2000,Callaway2000,Barabasi2000}.
The representation of complex systems has been limited to single networks
for a long time~\cite{Newman2010}. In many cases, however, coupling between
several networks takes place~\cite{Kivela2013,non2014}.
An important case is that of interdependency~\cite{Buldyrev2010,Li2012}
where there are two kinds of links: connectivity and dependency links.
An example of interdependent networks is the ensemble of the Internet and
the power supply grid where telecommunication is used to control power
plants and electric power is needed to supply communication
devices~\cite{Buldyrev2010}.
Connectivity links model the relation of the entities within the same
sector, spanning in the above example a power supply network and a 
telecommunication network.
Dependency links depict the basic supplies an entity
depends on which are supplied by entities in the other network. If a
supplier fails its dependent nodes fail as well. The system is viable if a
giant component of interconnected units exists in both networks.
In the 28 September 2003 blackout in Italy it came to evidence that the
interdependency of the two networks makes them more vulnerable than ever
thought before~\citep{Buldyrev2010}.
Similar relations occur in the economics between banks and firms or funds.
Banks are related through interbank loans, firms through supply chains and
the interdependence comes from loans and securities. Inappropriate asset
proportions can also lead to global avalanches as seen in the subprime
mortgage crisis~\cite{May2011}.

Interconnecting similar subsystems used to increase capacity was shown
beneficial as long as it does not open pathways to
cascades~\cite{Brummitt2012}.
However, in interdependent networks, the aspect of robustness was
considered with the conclusion that broadening the degree distribution of
the initial networks enhances vulnerability~\cite{Buldyrev2012}.
A cost-intensive intervention to strengthen robustness is to upgrade nodes
to be autonomous on some resources \cite{Schneider2013}.

\begin{figure}[tb]
\includegraphics{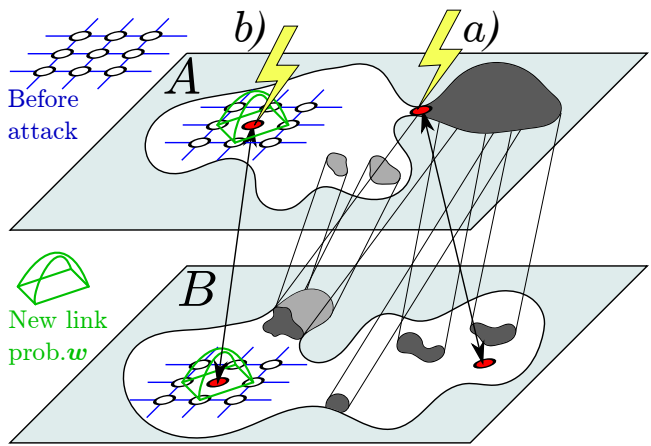}
\caption{\label{fig:cascade_schema} \emph{a)}
Failures, represented by red dots, affect the nodes one by one
in a random order. Whenever a node fails, its counterpart, that is,
the node in the other network which depends on it, fails as well.
In both networks, only the largest connected component (LCC) survives.
This constraint can cause further nodes to fail in both networks,
which trigger further shrinking of the LCC, and so on, illustrated by
the shaded areas.
\emph{b)} The neighbors of a failing node try to heal the network,
such that two functioning neighbors of a removed node establish a
connectivity link with probability $w$.}
\end{figure}

Because failures propagate rapidly in infrastructure networks,
they cannot be stopped by installing 
backup devices during the spreading of the damage.
but rather they require already existing systems. 
After the cascade of failures, damaged devices or elements can be replaced 
by new, functioning ones \emph{identical} to the originals~\cite{Havlin2012}.
In contrast to engineered systems, social or economic networks are
highly responsive and may react quickly~\cite{Onnela2010,Schweitzer2009}.
When a failure occurs considerable effort is made to reorganize the network
and rearrange the load of failing elements among functioning ones. The role
of the failing entities is taken over by \emph{similar} participants.
Such processes can be modeled by healing, i.e., substituting some of
the failed elements by new ones.
The timescale of an economic crisis is wide enough for the network to
completely restructure itself~\cite{Schweitzer2009}.
So far such mechanisms have only been studied for simple
networks~\cite{Wang5,Kenett2013,Havlin2012challenges}.
Here we extend the original model~\cite{Buldyrev2010} of
cascading failures of interdependent networks. After each removal, the
healing process attempts to bypass the removed node with a new
connectivity link (see Fig.~\ref{fig:cascade_schema}).
In this paper, we demonstrate how healing acts on
interdependent networks.

The outline of the paper is as follows.
In Sec.~\ref{sec:model} we define the node failure process in a dynamic way.
We introduce initial failures one by one to be able to apply healing at
every failure event.
Then we relate the original version of cascading failures to our model
as a special case and give formulas for comparing the order parameter of
the two models. The scaling properties of the healing are explained along
with the numeric results in Sec.~\ref{sec:scaling}.
In Sec.~\ref{sec:cascades}
we discuss the properties of the cascades with microscopic insight to the
model. Finally we conclude our findings in Sec.~\ref{sec:conclusion}.

\section{The model}
\label{sec:model}

In the standard model of interdependent networks~\cite{Li2012}
the computer-generated model-system is built up of
two topologically identical networks $A$ and $B$,
e.g., square lattices of size $N=L\times L$, where each node has
\emph{connectivity} links within the same network.
In addition, \emph{dependency} links couple between the networks,
which are bidirectional one-to-one relationships connecting randomly
selected pairs of nodes from the two networks. If any of the nodes fails
its dependent pair fails too.
A node in any network can function only if it is connected to the largest
connected component of that network  the node which it depends on is also
functional, otherwise it fails, i.e., it is removed from the network.

The existence of a macroscopic connected component in a single network is
treated by percolation theory. In the usual case, for a lattice it describes
a second-order phase transition between the phases with and without the
existence of a giant component~\cite{Stauffer1994}.
Adding interdependency allows cascades of failures to
propagate between the two networks. The threshold the network can survive
without collapse decreases considerably in this setting~\cite{Li2012}.

The collapse due to cascades was shown to be a first order transition
if the dependency links have unlimited range while the transition is of
second order if the range is less than a critical length
$r_c$~\cite{Li2012, Danziger2013}. Moreover,
the first order transition has a hybrid character with scaling on one
of its sides~\citep{Havlin2011,Baxter2012}.

\begin{figure}[tb]
\includegraphics{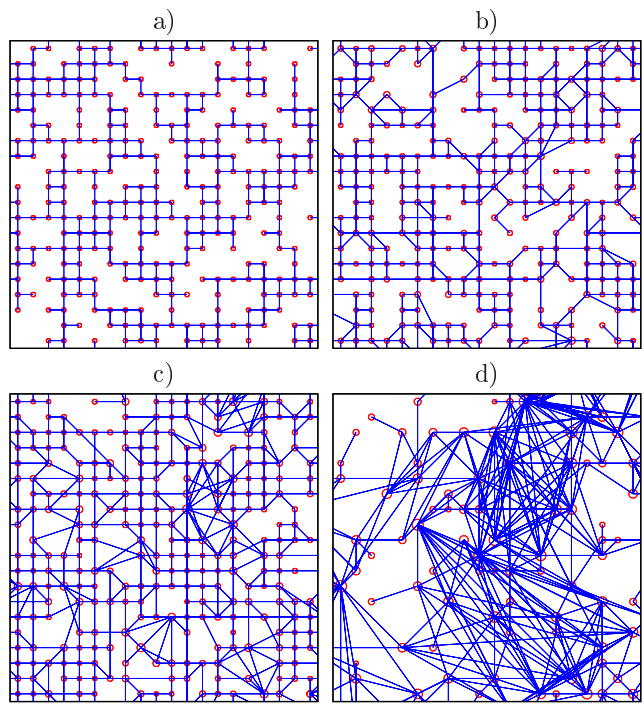}
\caption{\label{fig:insight} Part of Network $A$ of a
simulated system at $p=0.7$ at \emph{a)} no healing ($w=0.0$)
\emph{b)} below the critical healing ($w=0.2$, the average degree stays
below $4$) and
\emph{c)} slightly above the critical healing ($w=0.4$).
\emph{d)} This latter $w=0.4$ system is also represented at $p=0.2$ where
one can observe that the nodes get more and more connected and the healing 
process establishes links between distant nodes.}
\end{figure}

As mentioned in the Introduction we first introduce a dynamic process on
the interdependent network model.
In the setting of two interdependent networks of general topology this
dynamic process consists of
the repetition of attacks and relaxations to a rest via cascades.
(See Fig.~\ref{fig:cascade_schema}.)
Let us suppose that failures affect the nodes one by one in a random order
which defines a timeline. One time step is identified with the external
attack of one node. Time is measured by the number of time steps normalized
by $N$ for systems of different sizes to be comparable: 
\[ \mathrm{elapsed~time} = 1 - p = \frac{\mathrm{number~of~time~steps}}{N} \]
The externally introduced failure in network $A$ may separate the largest
connected component (LCC) into two or more parts where only the largest one
survives. 
All the failed nodes have dependency connections to nodes of the network $B$
causing their failure. Again, the LCC of $B$ may get fragmented and only the
largest part survives. This cascading procedure is repeated until no more
failures happen.
Of course, our model can easily be generalized to any number of
interdependent networks and any density of dependency links.

Our aim is to introduce healing into this dynamic model. The procedure is
as follows:
After an externally introduced failure (which may cut off a part of the LCC)
the healing step follows. Two remaining, functioning neighbors of a removed
node establish a connectivity link with an independent probability $w$.
(See part \emph{b)} in Fig.~\ref{fig:cascade_schema}.)
Then the dependent nodes of the removed nodes are removed from the other
network. After the propagation of the failure there, again, two functioning
neighbors of a removed node establish a connectivity link probability $w$.
Due to the separation of small components, further damages might propagate
back and forth within the network, always followed by a healing step.
Here, the healing step means that all pairs of neighbors of
each failed node is considered as a candidate for a new connectivity link
with an independent probability $w$, then, after having selected
the candidates, the connectivity links are established simultaneously.
The process goes on until no more separation of components occurs.
The healing links may change the topology considerably, bridging larger
and larger distances as the time goes on (Fig.~\ref{fig:insight}).
Once a critical fraction $(1-p_c)$ of nodes are removed, a catastrophic
cascade destroys the remaining system.

\begin{figure*}[t]
\includegraphics{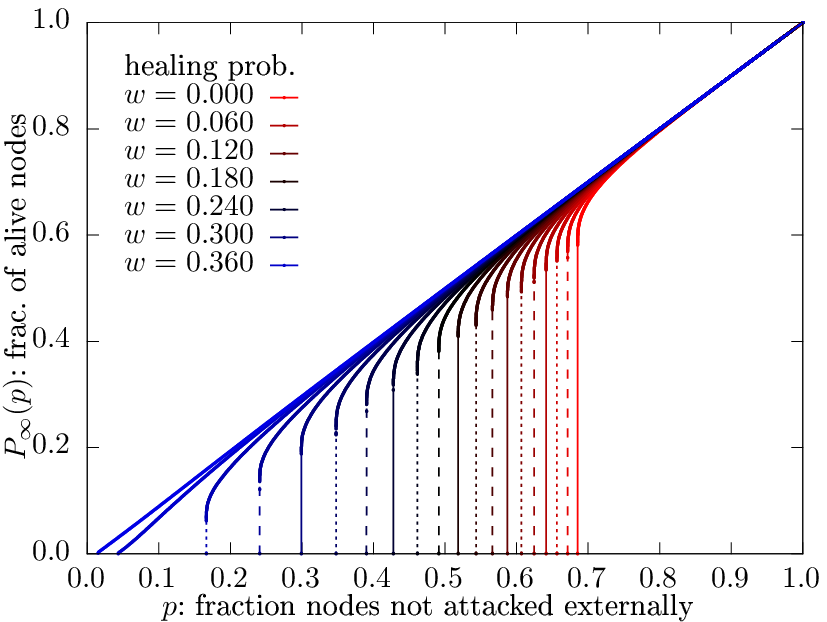}\hfill
\includegraphics{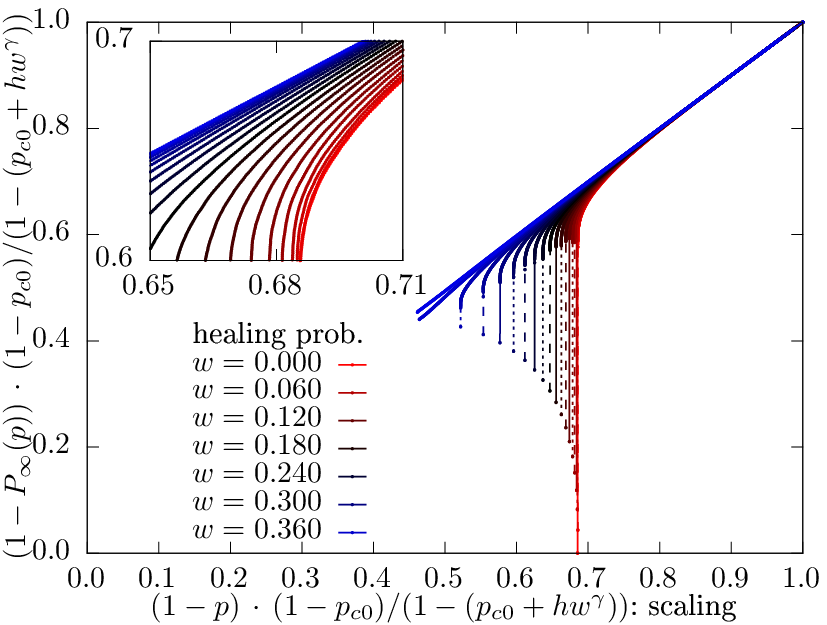}
\caption{\label{fig:orderparam} \emph{(left)}
The fraction $P_\infty(p)$ of remaining nodes of the original
$N=320\times320$ nodes as a function of the fraction $p$ of
nodes not attacked externally. Note: In order to sharply mark the
breakdown, averaging in variable $p$ is done for a given $P_\infty(p)$
over $60$ simulations. \emph{(right)} The same curves scaled on
each other using relation \eqref{powerlaw}. In \emph{(both)} parts,
plots from the right to the left correspond to the range $w=0.00$
to $w=0.38$ respectively with a step size $0.02$, solid lines indicate
steps of $0.06$.}
\end{figure*}

The $w=0$ case is simply the dynamic version of the well studied model of
Li \emph{et al.}. In \cite{Li2012} a fraction $(1-q)$ of the original network
is destroyed in the first step then the size of the giant component after
the relaxation of cascades is traced as a function of $q$. The important
difference between this procedure and ours is that in the version of
Li \emph{et al.} nodes may be accidentally attacked, which already fail
in our step-by-step (dynamic) model. Let $P_\infty$ denote the fraction of
remaining nodes as a function of the fraction of attacked nodes $(1-p)$
in the step-by-step model. The number of unattacked but disconnected nodes is
$[p-P_\infty(p)]N$. The probability of randomly destroying an already
disconnected (but not attacked) node is $(p-P_\infty)/p$, so the implicit
relation between the two attacking methods is 
\footnote{The integral \eqref{corresponding_p} can be numerically evaluated
more accurately by \unexpanded{$1-p(q) = \int_{P_\infty(q)}^1 \left( 1 -
\frac{p(\widetilde{P}_\infty)-\widetilde{P}_\infty}{p(\widetilde{P}_\infty)}
\right) p'(\widetilde{P}_\infty) \,\mathrm{d}\widetilde{P}_\infty$} where
\unexpanded{$p(P_\infty)$} is the inverse function of
\unexpanded{$P_\infty(p)$}.}
\begin{equation}\label{corresponding_p} 1-p(q) = \int_q^1 1 -
\frac{\widetilde{p}-P_\infty(\widetilde{p})}{\widetilde{p}}
\,\mathrm{d}\widetilde{p}.
\end{equation}
Due to the small false target ratio in the random attack, the threshold
values of the two models are close.
The extrapolated threshold value for the infinite system size in case $w=0$
is $p_c = 0.690\pm0.001$, in good agreement with the result of
Li \emph{et al.}.

\section{Scaling with the healing probability}
\label{sec:scaling}

The order parameter $P_\infty$ of our model depends not only on
the fraction of attacked nodes but also on the healing parameter.
According to one's intuition, the data show that the critical attack
$(1-p_c)$ increases monotonically with $w$. 

We executed Monte Carlo simulations of our model with both periodic and
open boundary conditions on square lattices starting networks of linear size
$L=20$, $40$, $80$, $160$ and $320$ with $960$, $480$, $240$, $120$ and $60$
runs respectively, and we measured that the execution time
in our implementation scaled approximately as $N^{2.3}=L^{4.6}$.
In the square lattices connectivity links join nodes to their nearest
neighbors within the same network. Dependency links were established by
first creating the trivial mapping between the topologically identical
lattices, then randomly shuffling the end of the links.
The $p_c$-s are then obtained averaging over the \emph{vertical} axis: for a
given number of surviving
nodes $P_\infty(p,w)$, we averaged the proportion of
nodes $1-p$ attacked one-by-one.
Fig.~\ref{fig:orderparam} shows the averaged curves for different values
of $w$.
The shape of the $P_\infty(p,w)$ curves suggests the scaling in the form of
anisotropic resizing from the $S(p=1,P_\infty=1)$ point:
\begin{equation} \label{scalingform}
  1-P_\infty(1-p,w)=1-a(w)\,P_\infty\left(\frac{1-p}{c(w)}, 0\right)
\end{equation}
which is asymptotically satisfied in the $w\to 0$ limit.

\begin{figure}[tb]
\includegraphics{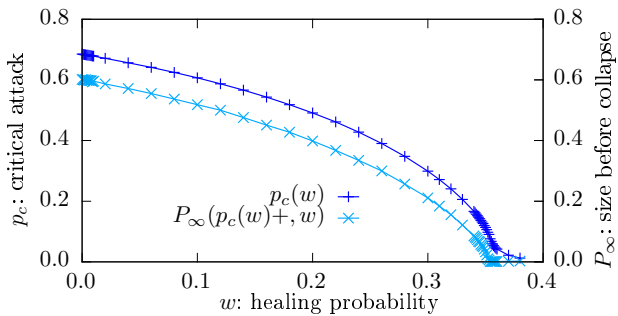}
\caption{\label{fig:criticalw}
$p_c$ as a function of $w$ depicts the fraction of unattacked nodes
at the transition for the starting $N=320\times320$ system size.
$P_\infty(p_c(w)+,w)$ is the giant component size just before the transition
as a function of $w$. Its non-zero value shows the jump in the first order transition and its zero value 
above $w_c = 0.351\pm0.002$ indicates a smooth transition.}
\end{figure}

\begin{figure}[bt]
\includegraphics{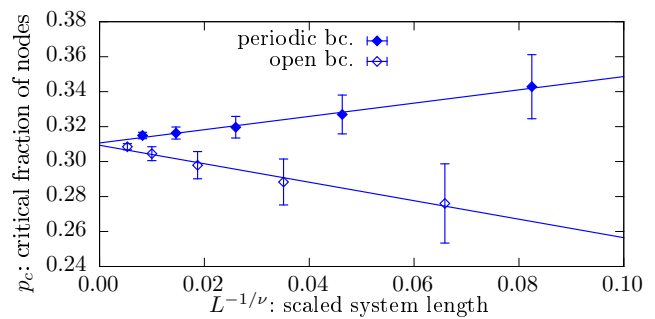}
\caption{\label{fig:fss} The standard deviation of the
critical attack fraction $1-p_c$ was used to obtain the length scaling
exponents $\nu_p=1.10\pm0.06$ and $\nu_o=1.20\pm0.06$ for periodic (filled)
and open (void symbols) boundary conditions as described in
Sec.~\ref{sec:scaling}. Then $p_c$ on the is plotted against $L^{-1/\nu}$
giving good collapse for the infinite system size.}
\end{figure}

In the infinite lattice limit, the initial few attacks almost surely occur
in different parts of the lattice and do not raise cascades, only the
attacked points fail, $P_\infty(p)=p$ if $p \sim 1$. The unit slope at
$S$ with respect to $p$ can be expressed by differentiation and yields
$a(w)\equiv c(w)$.
Let us express the fraction of unattacked nodes relative to the threshold
without healing: $\Delta p=p-p_{c0}\leq0$. The change in the threshold value
$\Delta p_c(w)=p_c(w)-p_{c0}$ can be identified by the largest $\Delta p$
where $P_\infty$ has an infinite slope (see Fig.~\ref{fig:criticalw}):
$\lim_{\Delta p\to \Delta p_c(w)+0} \frac{\partial}{\partial \Delta p}
 P_\infty(1-p_{c0}-\Delta p,w) = \infty$. Substituting it into
\eqref{scalingform} yields $a(w)=(1-p_{c0}-\Delta p_c(w))/(1-p_{c0})$.
The increase in lifetime, $-\Delta p_c(w)$, has a general scaling behavior
expressed in
\begin{equation} \label{powerlaw}
-\Delta p(w)= h\,w^\gamma
\end{equation}
for small $w$-s, in the range $[0.000,0.050]$. For the purpose of precise
measurement we created simulation data for all system sizes with step size
$0.001$ for $w\in[0.000,0.010]$ additional to that shown in
Fig.~\ref{fig:orderparam}.
The measurement is hampered by large fluctuations of the small systems,
therefore we extrapolated to infinite system size using standard
finite size scaling~\cite{Privman1990}. We used both systems with
periodic and open boundary conditions and measured the finite size
fluctuations in $p_c$ (approaching from the $p>p_c$ domain in accord with
the hybrid character of the transition) which yields slightly different
scaling exponents
within the error tolerance for the two systems (respectively
$\nu_p=1.10\pm0.06$ and $\nu_o=1.20\pm0.06$) from which we deduce
$\nu\approx1.15$.
The finite size scaling measurements, yielding $p_c = 0.690\pm0.001$, are
represented in Fig.~\ref{fig:fss}.

The parameters of Eq.~\eqref{powerlaw} are first fitted
for each system size $N=L\times L=20^2, 40^2, 80^2, 160^2 \text{ and } 320^2$,
then the infinite size limit is obtained using $1/L$ extrapolation.
The systems with periodic and open boundary conditions simulated at 
different system sizes collapse well yielding $h=0.703\pm0.005$ and
$\gamma=1.034\pm0.009$ for the infinite size network.

\begin{figure}[tb]
\includegraphics{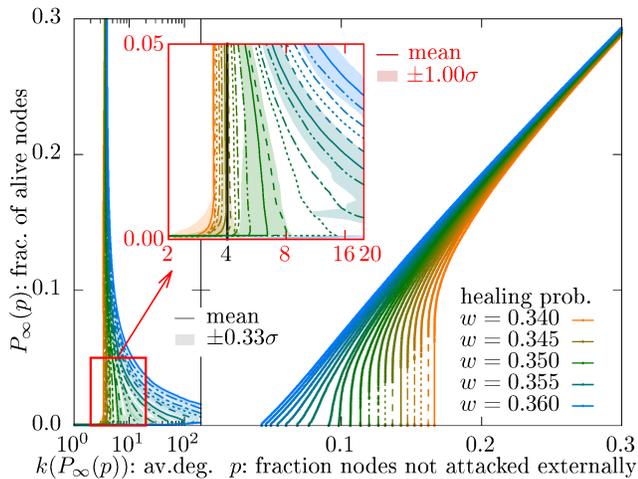}
\caption{\label{fig:degreechange} 
\emph{(left and inset)} The average degree on the horizontal axis as a
function of the fraction of dead nodes on the vertical axis for the
starting $N=320\times320$ system size. The average
degree remains constant for $w_c=0.348\pm0.003$. Plotted lines from the
right to the left correspond to the range $w=0.340$ to $w=0.360$
respectively with a step size $0.001$, solid lines indicate steps
of $0.005$. The shaded areas represent $0.33$ standard deviations in the
left part. In the inset, shaded areas are only plotted for solid lines and
represent $1.00$ standard deviation.
\emph{(right)} The fraction $P_\infty(p)$ of failing nodes 
as a function of the fraction $p$ of nodes not attacked externally using
the same averaging as in Fig.~\ref{fig:orderparam}.
Above $w_c=0.351\pm0.002$ there is no breakdown.}
\end{figure}

\section{Cascades change topology}
\label{sec:cascades}

We call cascades all events involving more nodes than the attacked one and
its dependency counterpart. The size (number of nodes involved compared to
the starting lattice size) of
typical cascades is small up to the point of breakdown. 

The healing dynamics changes the network topology and the average degree
as well. Fig.~\ref{fig:degreechange} allows us to describe a transition:
below a critical healing threshold $w_c$ we find a sharp breakdown in the
number of surviving nodes. The critical healing is defined as the lowest
$w$ for which the $P_\infty(p)$ function does not have an infinite slope.
In our simulation we observe $w_c = 0.351\pm0.002$.
For $w > w_c$ also there is no macroscopic cascade and $P_\infty(p)$
goes smoothly to zero in a second-order transition as $p$ decreases
(see also Fig.~\ref{fig:criticalw}).

The healing performed by the $k$ neighbors introduces $w\binom{k}{2}$ new
links on average.
A rough mean-field estimate of $w_c$ is the healing probability, which
conserves the average degree in the initial settings, leading to
$2w_c\binom{k}{2}=k$ (each link joins $2$ nodes). As the square lattice has
$k = 4$, the result is $w_c^{\text{mean-field}} = 1/3$. According to the
left plot in Fig.~\ref{fig:degreechange} we find that the
average degree $k = 4$ changes least through the simulation for
$w_c = 0.348\pm0.004$, which agrees well
with the critical healing determined from the $P_\infty$
curves~\footnote{In triangle lattice we measure $w_c\approx 0.265\pm0.005$,
and the heuristic argument gives $w_c=1/5$.}.
The change in the topology along with the trend of the average degree
can be observed in Fig.~\ref{fig:insight}. Below the critical healing $w_c$
the average degree is monotonically decreasing function of $1-P_\infty$ and
the connectivity links remain local, conserving the disordered lattice-like
topology. Thorough inspection
shows that all simulations end with a cascade wiping out all of the
remaining network at $p_c(w)$.
Above $w_c$ the healing promotes the formation of densely connected regions
and connectivity links begin to join distant nodes.
We remark that in the terminal stage the defined dynamics removes all nodes
and links in both cases. In summary, the difference is that  for $w<w_c$ the
process terminates with a macroscopic cascade, while for $w>w_c$ there is
no macroscopic cascade. In the latter case the average degree increases
until it has to decrease due to the small number of remaining nodes.

\section{Conclusions}
\label{sec:conclusion}

We examined the consequences of healing by edge formation in interdependent
networks under random attack. We found that the increase in resilience of
the network, measured in the number of survived attacks, has power-law
scaling with the probability $w$ of healing. By establishing new random
links in the neighborhood of the failed nodes, we delayed the collapse of
the network through the hindering of cascades.
We found that it is possible to completely suppress macroscopic
cascading failures for healing probabilities higher than a critical value
$w_c$; we demonstrated that this critical healing probability keeps the
average degree of the nodes close to the initial value while the network
topology changes.
By analyzing healing efficiency, these findings can aid in the development
of intervention strategies for crisis situations. The presented model
contains a number of unrealistic features, like the starting lattice, the
unbounded range and the high density of dependency links and the
non-locality of the healing links. Further studies should clarify the role
of these simplifications.

\section{Acknowledgements}
\label{sec:acknowledgements}
This work was partially supported by the
European Union and the European Social Fund through project FuturICT.hu
(Grant No.: TAMOP-4.2.2.C-11/1/KONV-2012-0013).
JK thanks MULTIPLEX, Grant No. 317532.
Thanks are due to \'Eva R\'acz for her help at the early stage of this work
and to Michael Danziger for a critical reading of the manuscript.


\begin{thebibliography}{25}%
\makeatletter
\providecommand \@ifxundefined [1]{%
 \@ifx{#1\undefined}
}%
\providecommand \@ifnum [1]{%
 \ifnum #1\expandafter \@firstoftwo
 \else \expandafter \@secondoftwo
 \fi
}%
\providecommand \@ifx [1]{%
 \ifx #1\expandafter \@firstoftwo
 \else \expandafter \@secondoftwo
 \fi
}%
\providecommand \natexlab [1]{#1}%
\providecommand \enquote  [1]{``#1''}%
\providecommand \bibnamefont  [1]{#1}%
\providecommand \bibfnamefont [1]{#1}%
\providecommand \citenamefont [1]{#1}%
\providecommand \href@noop [0]{\@secondoftwo}%
\providecommand \href [0]{\begingroup \@sanitize@url \@href}%
\providecommand \@href[1]{\@@startlink{#1}\@@href}%
\providecommand \@@href[1]{\endgroup#1\@@endlink}%
\providecommand \@sanitize@url [0]{\catcode `\\12\catcode `\$12\catcode
  `\&12\catcode `\#12\catcode `\^12\catcode `\_12\catcode `\%12\relax}%
\providecommand \@@startlink[1]{}%
\providecommand \@@endlink[0]{}%
\providecommand \url  [0]{\begingroup\@sanitize@url \@url }%
\providecommand \@url [1]{\endgroup\@href {#1}{\urlprefix }}%
\providecommand \urlprefix  [0]{URL }%
\providecommand \Eprint [0]{\href }%
\providecommand \doibase [0]{http://dx.doi.org/}%
\providecommand \selectlanguage [0]{\@gobble}%
\providecommand \bibinfo  [0]{\@secondoftwo}%
\providecommand \bibfield  [0]{\@secondoftwo}%
\providecommand \translation [1]{[#1]}%
\providecommand \BibitemOpen [0]{}%
\providecommand \bibitemStop [0]{}%
\providecommand \bibitemNoStop [0]{.\EOS\space}%
\providecommand \EOS [0]{\spacefactor3000\relax}%
\providecommand \BibitemShut  [1]{\csname bibitem#1\endcsname}%
\let\auto@bib@innerbib\@empty
\bibitem [{\citenamefont {Cohen}\ \emph {et~al.}(2000)\citenamefont {Cohen},
  \citenamefont {Erez}, \citenamefont {ben Avraham},\ and\ \citenamefont
  {Havlin}}]{Cohen2000}%
  \BibitemOpen
  \bibfield  {author} {\bibinfo {author} {\bibfnamefont {R.}~\bibnamefont
  {Cohen}}, \bibinfo {author} {\bibfnamefont {K.}~\bibnamefont {Erez}},
  \bibinfo {author} {\bibfnamefont {D.}~\bibnamefont {ben Avraham}}, \ and\
  \bibinfo {author} {\bibfnamefont {S.}~\bibnamefont {Havlin}},\ }\href
  {\doibase 10.1103/PhysRevLett.85.4626} {\bibfield  {journal} {\bibinfo
  {journal} {Phys. Rev. Lett.}\ }\textbf {\bibinfo {volume} {85}},\ \bibinfo
  {pages} {4626} (\bibinfo {year} {2000})}\BibitemShut {NoStop}%
\bibitem [{\citenamefont {Callaway}\ \emph {et~al.}(2000)\citenamefont
  {Callaway}, \citenamefont {Newman}, \citenamefont {Strogatz},\ and\
  \citenamefont {Watts}}]{Callaway2000}%
  \BibitemOpen
  \bibfield  {author} {\bibinfo {author} {\bibfnamefont {D.~S.}\ \bibnamefont
  {Callaway}}, \bibinfo {author} {\bibfnamefont {M.~E.~J.}\ \bibnamefont
  {Newman}}, \bibinfo {author} {\bibfnamefont {S.~H.}\ \bibnamefont
  {Strogatz}}, \ and\ \bibinfo {author} {\bibfnamefont {D.~J.}\ \bibnamefont
  {Watts}},\ }\href {\doibase 10.1103/PhysRevLett.85.5468} {\bibfield
  {journal} {\bibinfo  {journal} {{Phys. Rev. Lett.}}\ }\textbf {\bibinfo
  {volume} {{85}}},\ \bibinfo {pages} {5468} (\bibinfo {year}
  {{2000}})}\BibitemShut {NoStop}%
\bibitem [{\citenamefont {Albert}\ \emph {et~al.}(2000)\citenamefont {Albert},
  \citenamefont {Jeong},\ and\ \citenamefont {Barabasi}}]{Barabasi2000}%
  \BibitemOpen
  \bibfield  {author} {\bibinfo {author} {\bibfnamefont {R.}~\bibnamefont
  {Albert}}, \bibinfo {author} {\bibfnamefont {H.}~\bibnamefont {Jeong}}, \
  and\ \bibinfo {author} {\bibfnamefont {A.~L.}\ \bibnamefont {Barabasi}},\
  }\href {\doibase 10.1038/35019019} {\bibfield  {journal} {\bibinfo  {journal}
  {{Nature}}\ }\textbf {\bibinfo {volume} {{406}}},\ \bibinfo {pages} {378}
  (\bibinfo {year} {{2000}})}\BibitemShut {NoStop}%
\bibitem [{\citenamefont {Newman}(2010)}]{Newman2010}%
  \BibitemOpen
  \bibfield  {author} {\bibinfo {author} {\bibfnamefont {M.~E.~J.}\
  \bibnamefont {Newman}},\ }\href {\doibase
  10.1093/acprof:oso/9780199206650.001.0001} {\emph {\bibinfo {title}
  {{Networks: An Introduction}}}}\ (\bibinfo  {publisher} {{Oxford University
  Press}},\ \bibinfo {year} {{2010}})\BibitemShut {NoStop}%
\bibitem [{\citenamefont {Kivel\"a}\ \emph {et~al.}(2013)\citenamefont
  {Kivel\"a}, \citenamefont {Arenas}, \citenamefont {Barthelemy}, \citenamefont
  {Gleeson}, \citenamefont {Moreno},\ and\ \citenamefont
  {Porter}}]{Kivela2013}%
  \BibitemOpen
  \bibfield  {author} {\bibinfo {author} {\bibfnamefont {M.}~\bibnamefont
  {Kivel\"a}}, \bibinfo {author} {\bibfnamefont {A.}~\bibnamefont {Arenas}},
  \bibinfo {author} {\bibfnamefont {M.}~\bibnamefont {Barthelemy}}, \bibinfo
  {author} {\bibfnamefont {J.~P.}\ \bibnamefont {Gleeson}}, \bibinfo {author}
  {\bibfnamefont {Y.}~\bibnamefont {Moreno}}, \ and\ \bibinfo {author}
  {\bibfnamefont {M.~A.}\ \bibnamefont {Porter}},\ }\href@noop {} {\bibfield
  {journal} {\bibinfo  {journal} {{pre-print}}\ } (\bibinfo {year} {{2013}})},\
  \Eprint {http://arxiv.org/abs/1309.7233} {arXiv:1309.7233 [{soc-ph}]}
  \BibitemShut {NoStop}%
\bibitem [{\citenamefont {Kenett}\ \emph {et~al.}(2014)\citenamefont {Kenett},
  \citenamefont {Gao}, \citenamefont {Huang}, \citenamefont {Shao},
  \citenamefont {Vodenska}, \citenamefont {Buldyrev}, \citenamefont {Paul},
  \citenamefont {Stanley},\ and\ \citenamefont {Havlin}}]{non2014}%
  \BibitemOpen
  \bibfield  {author} {\bibinfo {author} {\bibfnamefont {D.~Y.}\ \bibnamefont
  {Kenett}}, \bibinfo {author} {\bibfnamefont {J.}~\bibnamefont {Gao}},
  \bibinfo {author} {\bibfnamefont {X.}~\bibnamefont {Huang}}, \bibinfo
  {author} {\bibfnamefont {S.}~\bibnamefont {Shao}}, \bibinfo {author}
  {\bibfnamefont {I.}~\bibnamefont {Vodenska}}, \bibinfo {author}
  {\bibfnamefont {S.~V.}\ \bibnamefont {Buldyrev}}, \bibinfo {author}
  {\bibfnamefont {G.}~\bibnamefont {Paul}}, \bibinfo {author} {\bibfnamefont
  {H.~E.}\ \bibnamefont {Stanley}}, \ and\ \bibinfo {author} {\bibfnamefont
  {S.}~\bibnamefont {Havlin}},\ }in\ \href {\doibase
  10.1007/978-3-319-03518-5_1} {\emph {\bibinfo {booktitle} {Networks of
  Networks: The Last Frontier of Complexity}}},\ \bibinfo {series and number}
  {Understanding Complex Systems},\ \bibinfo {editor} {edited by\ \bibinfo
  {editor} {\bibfnamefont {G.}~\bibnamefont {D'Agostino}}\ and\ \bibinfo
  {editor} {\bibfnamefont {A.}~\bibnamefont {Scala}}}\ (\bibinfo  {publisher}
  {Springer International Publishing},\ \bibinfo {year} {2014})\ pp.\ \bibinfo
  {pages} {3--36}\BibitemShut {NoStop}%
\bibitem [{\citenamefont {Buldyrev}\ \emph {et~al.}(2010)\citenamefont
  {Buldyrev}, \citenamefont {Parshani}, \citenamefont {Paul}, \citenamefont
  {Stanley},\ and\ \citenamefont {Havlin}}]{Buldyrev2010}%
  \BibitemOpen
  \bibfield  {author} {\bibinfo {author} {\bibfnamefont {S.~V.}\ \bibnamefont
  {Buldyrev}}, \bibinfo {author} {\bibfnamefont {R.}~\bibnamefont {Parshani}},
  \bibinfo {author} {\bibfnamefont {G.}~\bibnamefont {Paul}}, \bibinfo {author}
  {\bibfnamefont {H.~E.}\ \bibnamefont {Stanley}}, \ and\ \bibinfo {author}
  {\bibfnamefont {S.}~\bibnamefont {Havlin}},\ }\href {\doibase
  10.1038/nature08932} {\bibfield  {journal} {\bibinfo  {journal} {{Nature}}\
  }\textbf {\bibinfo {volume} {{464}}},\ \bibinfo {pages} {1025} (\bibinfo
  {year} {{2010}})}\BibitemShut {NoStop}%
\bibitem [{\citenamefont {Li}\ \emph {et~al.}(2012)\citenamefont {Li},
  \citenamefont {Bashan}, \citenamefont {Buldyrev}, \citenamefont {Stanley},\
  and\ \citenamefont {Havlin}}]{Li2012}%
  \BibitemOpen
  \bibfield  {author} {\bibinfo {author} {\bibfnamefont {W.}~\bibnamefont
  {Li}}, \bibinfo {author} {\bibfnamefont {A.}~\bibnamefont {Bashan}}, \bibinfo
  {author} {\bibfnamefont {S.~V.}\ \bibnamefont {Buldyrev}}, \bibinfo {author}
  {\bibfnamefont {H.~E.}\ \bibnamefont {Stanley}}, \ and\ \bibinfo {author}
  {\bibfnamefont {S.}~\bibnamefont {Havlin}},\ }\href {\doibase
  10.1103/PhysRevLett.108.228702} {\bibfield  {journal} {\bibinfo  {journal}
  {{Phys. Rev. Lett.}}\ }\textbf {\bibinfo {volume} {{108}}},\ \bibinfo {pages}
  {228702} (\bibinfo {year} {{2012}})}\BibitemShut {NoStop}%
\bibitem [{\citenamefont {Haldane}\ and\ \citenamefont {May}(2011)}]{May2011}%
  \BibitemOpen
  \bibfield  {author} {\bibinfo {author} {\bibfnamefont {A.~G.}\ \bibnamefont
  {Haldane}}\ and\ \bibinfo {author} {\bibfnamefont {R.~M.}\ \bibnamefont
  {May}},\ }\href {\doibase 10.1038/nature09659} {\bibfield  {journal}
  {\bibinfo  {journal} {{Nature}}\ }\textbf {\bibinfo {volume} {{469}}},\
  \bibinfo {pages} {351} (\bibinfo {year} {{2011}})}\BibitemShut {NoStop}%
\bibitem [{\citenamefont {Brummitt}\ \emph {et~al.}(2012)\citenamefont
  {Brummitt}, \citenamefont {D'Souza},\ and\ \citenamefont
  {Leicht}}]{Brummitt2012}%
  \BibitemOpen
  \bibfield  {author} {\bibinfo {author} {\bibfnamefont {C.~D.}\ \bibnamefont
  {Brummitt}}, \bibinfo {author} {\bibfnamefont {R.~M.}\ \bibnamefont
  {D'Souza}}, \ and\ \bibinfo {author} {\bibfnamefont {E.~A.}\ \bibnamefont
  {Leicht}},\ }\href {\doibase 10.1073/pnas.1110586109} {\bibfield  {journal}
  {\bibinfo  {journal} {{Proc. Nat. Ac. Sci. of the USA}}\ }\textbf {\bibinfo
  {volume} {{109}}},\ \bibinfo {pages} {E680} (\bibinfo {year}
  {{2012}})}\BibitemShut {NoStop}%
\bibitem [{\citenamefont {Gao}\ \emph {et~al.}(2012{\natexlab{a}})\citenamefont
  {Gao}, \citenamefont {Buldyrev}, \citenamefont {Havlin},\ and\ \citenamefont
  {Stanley}}]{Buldyrev2012}%
  \BibitemOpen
  \bibfield  {author} {\bibinfo {author} {\bibfnamefont {J.}~\bibnamefont
  {Gao}}, \bibinfo {author} {\bibfnamefont {S.~V.}\ \bibnamefont {Buldyrev}},
  \bibinfo {author} {\bibfnamefont {S.}~\bibnamefont {Havlin}}, \ and\ \bibinfo
  {author} {\bibfnamefont {H.~E.}\ \bibnamefont {Stanley}},\ }\href {\doibase
  10.1103/PhysRevE.85.066134} {\bibfield  {journal} {\bibinfo  {journal}
  {{Phys. Rev. E}}\ }\textbf {\bibinfo {volume} {{85}}},\ \bibinfo {pages}
  {066134} (\bibinfo {year} {{2012}}{\natexlab{a}})}\BibitemShut {NoStop}%
\bibitem [{\citenamefont {Schneider}\ \emph {et~al.}(2013)\citenamefont
  {Schneider}, \citenamefont {Yazdani}, \citenamefont {Araujo}, \citenamefont
  {Havlin},\ and\ \citenamefont {Herrmann}}]{Schneider2013}%
  \BibitemOpen
  \bibfield  {author} {\bibinfo {author} {\bibfnamefont {C.~M.}\ \bibnamefont
  {Schneider}}, \bibinfo {author} {\bibfnamefont {N.}~\bibnamefont {Yazdani}},
  \bibinfo {author} {\bibfnamefont {N.~A.~M.}\ \bibnamefont {Araujo}}, \bibinfo
  {author} {\bibfnamefont {S.}~\bibnamefont {Havlin}}, \ and\ \bibinfo {author}
  {\bibfnamefont {H.~J.}\ \bibnamefont {Herrmann}},\ }\href {\doibase
  10.1038/srep01969} {\bibfield  {journal} {\bibinfo  {journal} {{Sci. Rep.}}\
  }\textbf {\bibinfo {volume} {{3}}},\ \bibinfo {pages} {1969} (\bibinfo {year}
  {{2013}})}\BibitemShut {NoStop}%
\bibitem [{\citenamefont {Gao}\ \emph {et~al.}(2012{\natexlab{b}})\citenamefont
  {Gao}, \citenamefont {Buldyrev}, \citenamefont {Stanley},\ and\ \citenamefont
  {Havlin}}]{Havlin2012}%
  \BibitemOpen
  \bibfield  {author} {\bibinfo {author} {\bibfnamefont {J.}~\bibnamefont
  {Gao}}, \bibinfo {author} {\bibfnamefont {S.~V.}\ \bibnamefont {Buldyrev}},
  \bibinfo {author} {\bibfnamefont {H.~E.}\ \bibnamefont {Stanley}}, \ and\
  \bibinfo {author} {\bibfnamefont {S.}~\bibnamefont {Havlin}},\ }\href
  {\doibase 10.1038/nphys2180} {\bibfield  {journal} {\bibinfo  {journal}
  {{Nat. Phys.}}\ }\textbf {\bibinfo {volume} {{8}}},\ \bibinfo {pages} {40}
  (\bibinfo {year} {{2012}}{\natexlab{b}})}\BibitemShut {NoStop}%
\bibitem [{\citenamefont {Mucha}\ \emph {et~al.}(2010)\citenamefont {Mucha},
  \citenamefont {Richardson}, \citenamefont {Macon}, \citenamefont {Porter},\
  and\ \citenamefont {Onnela}}]{Onnela2010}%
  \BibitemOpen
  \bibfield  {author} {\bibinfo {author} {\bibfnamefont {P.~J.}\ \bibnamefont
  {Mucha}}, \bibinfo {author} {\bibfnamefont {T.}~\bibnamefont {Richardson}},
  \bibinfo {author} {\bibfnamefont {K.}~\bibnamefont {Macon}}, \bibinfo
  {author} {\bibfnamefont {M.~A.}\ \bibnamefont {Porter}}, \ and\ \bibinfo
  {author} {\bibfnamefont {J.-P.}\ \bibnamefont {Onnela}},\ }\href {\doibase
  10.1126/science.1184819} {\bibfield  {journal} {\bibinfo  {journal}
  {{Science}}\ }\textbf {\bibinfo {volume} {{328}}},\ \bibinfo {pages} {876}
  (\bibinfo {year} {{2010}})}\BibitemShut {NoStop}%
\bibitem [{\citenamefont {Schweitzer}\ \emph {et~al.}(2009)\citenamefont
  {Schweitzer}, \citenamefont {Fagiolo}, \citenamefont {Sornette},
  \citenamefont {Vega-Redondo}, \citenamefont {Vespignani},\ and\ \citenamefont
  {White}}]{Schweitzer2009}%
  \BibitemOpen
  \bibfield  {author} {\bibinfo {author} {\bibfnamefont {F.}~\bibnamefont
  {Schweitzer}}, \bibinfo {author} {\bibfnamefont {G.}~\bibnamefont {Fagiolo}},
  \bibinfo {author} {\bibfnamefont {D.}~\bibnamefont {Sornette}}, \bibinfo
  {author} {\bibfnamefont {F.}~\bibnamefont {Vega-Redondo}}, \bibinfo {author}
  {\bibfnamefont {A.}~\bibnamefont {Vespignani}}, \ and\ \bibinfo {author}
  {\bibfnamefont {D.~R.}\ \bibnamefont {White}},\ }\href {\doibase
  10.1126/science.1173644} {\bibfield  {journal} {\bibinfo  {journal}
  {{Science}}\ }\textbf {\bibinfo {volume} {{325}}},\ \bibinfo {pages} {422}
  (\bibinfo {year} {{2009}})}\BibitemShut {NoStop}%
\bibitem [{\citenamefont {Wang}(2013)}]{Wang5}%
  \BibitemOpen
  \bibfield  {author} {\bibinfo {author} {\bibfnamefont {J.}~\bibnamefont
  {Wang}},\ }\href {\doibase 10.1016/j.physa.2013.01.013} {\bibfield  {journal}
  {\bibinfo  {journal} {{Physica A-Stat. Mech. and Appl.}}\ }\textbf {\bibinfo
  {volume} {{392}}},\ \bibinfo {pages} {2257} (\bibinfo {year}
  {{2013}})}\BibitemShut {NoStop}%
\bibitem [{\citenamefont {Majdandzic}\ \emph {et~al.}(2014)\citenamefont
  {Majdandzic}, \citenamefont {Podobnik}, \citenamefont {Buldyrev},
  \citenamefont {Kenett}, \citenamefont {Havlin},\ and\ \citenamefont
  {Stanley}}]{Kenett2013}%
  \BibitemOpen
  \bibfield  {author} {\bibinfo {author} {\bibfnamefont {A.}~\bibnamefont
  {Majdandzic}}, \bibinfo {author} {\bibfnamefont {B.}~\bibnamefont
  {Podobnik}}, \bibinfo {author} {\bibfnamefont {S.~V.}\ \bibnamefont
  {Buldyrev}}, \bibinfo {author} {\bibfnamefont {D.~Y.}\ \bibnamefont
  {Kenett}}, \bibinfo {author} {\bibfnamefont {S.}~\bibnamefont {Havlin}}, \
  and\ \bibinfo {author} {\bibfnamefont {H.~E.}\ \bibnamefont {Stanley}},\
  }\href {\doibase 10.1038/nphys2819} {\bibfield  {journal} {\bibinfo
  {journal} {{Nat. Phys.}}\ }\textbf {\bibinfo {volume} {{10}}},\ \bibinfo
  {pages} {34} (\bibinfo {year} {{2014}})}\BibitemShut {NoStop}%
\bibitem [{\citenamefont {Havlin}\ \emph {et~al.}(2012)\citenamefont {Havlin},
  \citenamefont {Kenett}, \citenamefont {Ben-Jacob}, \citenamefont {Bunde},
  \citenamefont {Cohen}, \citenamefont {Hermann}, \citenamefont {Kantelhardt},
  \citenamefont {Kert\'esz}, \citenamefont {Kirkpatrick}, \citenamefont
  {Kurths}, \citenamefont {Portugali},\ and\ \citenamefont
  {Solomon}}]{Havlin2012challenges}%
  \BibitemOpen
  \bibfield  {author} {\bibinfo {author} {\bibfnamefont {S.}~\bibnamefont
  {Havlin}}, \bibinfo {author} {\bibfnamefont {D.~Y.}\ \bibnamefont {Kenett}},
  \bibinfo {author} {\bibfnamefont {E.}~\bibnamefont {Ben-Jacob}}, \bibinfo
  {author} {\bibfnamefont {A.}~\bibnamefont {Bunde}}, \bibinfo {author}
  {\bibfnamefont {R.}~\bibnamefont {Cohen}}, \bibinfo {author} {\bibfnamefont
  {H.}~\bibnamefont {Hermann}}, \bibinfo {author} {\bibfnamefont
  {J.}~\bibnamefont {Kantelhardt}}, \bibinfo {author} {\bibfnamefont
  {J.}~\bibnamefont {Kert\'esz}}, \bibinfo {author} {\bibfnamefont
  {S.}~\bibnamefont {Kirkpatrick}}, \bibinfo {author} {\bibfnamefont
  {J.}~\bibnamefont {Kurths}}, \bibinfo {author} {\bibfnamefont
  {J.}~\bibnamefont {Portugali}}, \ and\ \bibinfo {author} {\bibfnamefont
  {S.}~\bibnamefont {Solomon}},\ }\href {\doibase 10.1140/epjst/e2012-01695-x}
  {\bibfield  {journal} {\bibinfo  {journal} {The European Physical Journal
  Special Topics}\ }\textbf {\bibinfo {volume} {214}},\ \bibinfo {pages} {273}
  (\bibinfo {year} {2012})}\BibitemShut {NoStop}%
\bibitem [{\citenamefont {Stauffer}\ and\ \citenamefont
  {Aharony}(1994)}]{Stauffer1994}%
  \BibitemOpen
  \bibfield  {author} {\bibinfo {author} {\bibfnamefont {D.}~\bibnamefont
  {Stauffer}}\ and\ \bibinfo {author} {\bibfnamefont {A.}~\bibnamefont
  {Aharony}},\ }\href {http://www.crcpress.com/product/isbn/9780748402533}
  {\emph {\bibinfo {title} {{Intorduction to Percolation Theory}}}},\ \bibinfo
  {edition} {{2nd}}\ ed.\ (\bibinfo  {publisher} {{Taylor and Francis,
  London}},\ \bibinfo {year} {{1994}})\BibitemShut {NoStop}%
\bibitem [{\citenamefont {Danziger}\ \emph {et~al.}(2013)\citenamefont
  {Danziger}, \citenamefont {Bashan}, \citenamefont {Berezin},\ and\
  \citenamefont {Havlin}}]{Danziger2013}%
  \BibitemOpen
  \bibfield  {author} {\bibinfo {author} {\bibfnamefont {M.}~\bibnamefont
  {Danziger}}, \bibinfo {author} {\bibfnamefont {A.}~\bibnamefont {Bashan}},
  \bibinfo {author} {\bibfnamefont {Y.}~\bibnamefont {Berezin}}, \ and\
  \bibinfo {author} {\bibfnamefont {S.}~\bibnamefont {Havlin}},\ }in\ \href
  {\doibase 10.1109/SITIS.2013.101} {\emph {\bibinfo {booktitle} {{2013
  International Conference on Signal-Image Technology and Internet-Based
  Systems (SITIS)}}}}\ (\bibinfo  {publisher} {IEEE},\ \bibinfo {year}
  {{2013}})\ pp.\ \bibinfo {pages} {619--625}\BibitemShut {NoStop}%
\bibitem [{\citenamefont {Hu}\ \emph {et~al.}(2011)\citenamefont {Hu},
  \citenamefont {Ksherim}, \citenamefont {Cohen},\ and\ \citenamefont
  {Havlin}}]{Havlin2011}%
  \BibitemOpen
  \bibfield  {author} {\bibinfo {author} {\bibfnamefont {Y.}~\bibnamefont
  {Hu}}, \bibinfo {author} {\bibfnamefont {B.}~\bibnamefont {Ksherim}},
  \bibinfo {author} {\bibfnamefont {R.}~\bibnamefont {Cohen}}, \ and\ \bibinfo
  {author} {\bibfnamefont {S.}~\bibnamefont {Havlin}},\ }\href {\doibase
  10.1103/PhysRevE.84.066116} {\bibfield  {journal} {\bibinfo  {journal}
  {{Phys. Rev. E}}\ }\textbf {\bibinfo {volume} {{84}}},\ \bibinfo {pages}
  {066116} (\bibinfo {year} {{2011}})}\BibitemShut {NoStop}%
\bibitem [{\citenamefont {Baxter}\ \emph {et~al.}(2012)\citenamefont {Baxter},
  \citenamefont {Dorogovtsev}, \citenamefont {Goltsev},\ and\ \citenamefont
  {Mendes}}]{Baxter2012}%
  \BibitemOpen
  \bibfield  {author} {\bibinfo {author} {\bibfnamefont {G.~J.}\ \bibnamefont
  {Baxter}}, \bibinfo {author} {\bibfnamefont {S.~N.}\ \bibnamefont
  {Dorogovtsev}}, \bibinfo {author} {\bibfnamefont {A.~V.}\ \bibnamefont
  {Goltsev}}, \ and\ \bibinfo {author} {\bibfnamefont {J.~F.~F.}\ \bibnamefont
  {Mendes}},\ }\href {\doibase 10.1103/PhysRevLett.109.248701} {\bibfield
  {journal} {\bibinfo  {journal} {{Phys. Rev. Lett.}}\ }\textbf {\bibinfo
  {volume} {{109}}},\ \bibinfo {pages} {248701} (\bibinfo {year}
  {{2012}})}\BibitemShut {NoStop}%
\bibitem [{Note1()}]{Note1}%
  \BibitemOpen
  \bibinfo {note} {The integral \protect \textup {\hbox {\mathsurround \z@
  \protect \normalfont (\ignorespaces \ref {corresponding_p}\unskip
  \@@italiccorr )}} can be numerically evaluated more accurately by $1-p(q) =
  \int _{P_\infty (q)}^1 \left ( 1 - \frac {p(\widetilde {P}_\infty
  )-\widetilde {P}_\infty }{p(\widetilde {P}_\infty )} \right ) p'(\widetilde
  {P}_\infty ) \,\mathrm {d}\widetilde {P}_\infty $ where $p(P_\infty )$ is the
  inverse function of $P_\infty (p)$.}\BibitemShut {Stop}%
\bibitem [{\citenamefont {Privman}(1990)}]{Privman1990}%
  \BibitemOpen
  \bibinfo {editor} {\bibfnamefont {V.}~\bibnamefont {Privman}},\ ed.,\ \href
  {http://www.worldscientific.com/worldscibooks/10.1142/1011} {\emph {\bibinfo
  {title} {Finite Size Scaling and Numerical Simulation of Statistical
  Systems}}}\ (\bibinfo  {publisher} {World Scientific Publishing Company,
  Inc.},\ \bibinfo {year} {1990})\BibitemShut {NoStop}%
\bibitem [{Note2()}]{Note2}%
  \BibitemOpen
  \bibinfo {note} {In triangle lattice we measure $w_c\approx 0.265\pm 0.005$,
  and the heuristic argument gives $w_c=1/5$.}\BibitemShut {Stop}%
\end{thebibliography}

%

\end{document}